\documentclass[aps,prb,twocolumn,showpacs]{revtex4-1}
\usepackage{graphicx}
\usepackage{bm,amsmath,amssymb}
\newcommand{\bos}{\boldsymbol}

\begin{document}
\title{The topological effect on the Mechanical properties of polymer knots}
\author{Yani Zhao$^1$}
\email{y.zhao@cent.uw.edu.pl}
\author{Franco Ferrari$^2$}
\email{ferrari@fermi.fiz.univ.szczecin.pl}
\affiliation{$^1$Centre of New Technologies, University of Warsaw,
  Warsaw, Poland\\
  $^2$CASA* and Institute of Physics, University of Szczecin,
  Szczecin, Poland} 
\date{\today}

\begin{abstract}
  The mechanical properties of polymer knots 
under stretching in a bad or good solvent
  are investigated by applying a
  given force $F$ to a point of the knot while keeping another point fixed.
  The Monte Carlo sampling of the polymer conformations on a simple
cubic lattice  is performed using a variant of the Wang-Landau
algorithm.
The results of the calculations of the specific energy,  specific heat
capacity and gyration radius for several knot topologies show a
general trend in the behavior of 
short polymer knots with lengths up to
seventy lattice units. At low tensile force $F$, knots can be
found either in a compact or an extended phase, depending if the
temperature is low or high. At any temperature, with
increasing values 
of the force $F$, a polymer knot undergoes a phase transition to a
stretched state. This transition is characterized by a strong peak in
the heat capacity. There is also a minor peak, which corresponds
to a transition occurring at
low temperatures when the conformations of polymers in the stretched phase  
become swollen with increasing temperatures. It is also shown
that the behavior of short polymer rings is strongly influenced by
topological effects.
The limitations in the number of accessible energy states due to
topological constraints is particularly evident in knots of small size and
such that  their minimum number of crossing according to the Rolfsen
knot table is high. An example is provided
by a cinquefoil knot $5_1$  with a length of only fifty lattice units.
The thermal and mechanical properties of knots
that can be represented 
with diagrams having 
the same minimum number of crossings, are very similar.
The size effects on the behavior of polymer knots have been analyzed too.
Surprisingly, it is found that topological effects fade out very fast with
increasing polymer length.
\end{abstract}
\maketitle
\section{Introduction}\label{introd}
Closed polymer rings with nontrivial topological properties are a
major pattern in nature. The formation of DNA knots is a phenomenon that
has been observed in all the three domains of life \cite{d1,d2}.
Bacterial DNA often occurs in the form of knots that 
are sometimes heavily linked together. Knots are present also in the
DNA of viruses. For instance, it has been shown that 95\%
of the DNA
molecules extracted 
from tailless mutants of phage P4 {\it vir 1 del22} are highly knotted
\cite{per95}.  Knots appear in proteins too, although 
not so frequently  as in  DNA \cite{d3}.
So far,  the following knots have been found in proteins:
the trefoil knot $3_1$, the figure-eight knot
$4_1$, the penta knot $5_2$ and the Stevedore knot $6_1$
\cite{61knot}.  It should be however recalled that the trajectories of
proteins are open, so that
their knots do not sustain the full topological properties of mathematical
knots.

The synthesis of  artificial polymer
knots and links has been realized more than twenty years ago and there
have been in this subject many new
developments \cite{Dietrich-Buchecker1989,schappacher,ohta2009,Ohta2012}.  
For example, the first artificial trefoil knot  has been synthesized
by Dietrich-Buchecker et al. \cite{Dietrich-Buchecker1989} in 1989,
while  knotted ring polystyrene with high molecular weight has been
successfully synthesized by intramolecular cyclization reactions in
poor solvents by Ohta et al. \cite{Ohta2012}.
The
presence of knots has several consequences 
on the behavior 
of polymer materials which, most important, are measurable in
experiments and may be used to test theoretical results, see for example
\cite{Arai1999,Saitta1999,levene,trigueiros,
s2,s3,elbaz,radloff,likos,electrophoresis2,nahum,k2008,Hossain}.
Very recently, the paper \cite{newpaper} appeared with an updated list of
advances and references. Topological effects in the glass transition
of polymer knots have been tested using the methods of calorimetry in
\cite{HossainPhD}. 

Analytically, more specifically using the methods of field theory and
renormalization group theory that have been so successful in the case
of linear polymer chains, up to now it is possible to study only linked
polymers rings, the so-called catenanes
\cite{FFIL2,FFAP,FFIL,FFIL3}. Despite several 
efforts, we  
cite here only the most recent ones \cite{ffmpyz,rohwer}, there is
still no analytical model for polymer knots.
For this reason, in this work the properties
of a single knot subjected to a tensile force
will be investigated 
numerically. For our purposes, it will be convenient to use
the particular variant \cite{newwl} of the original
Wang-Landau Monte Carlo algorithm \cite{wl}. Both the original
algorithm and the variant have been
already  
applied to  polymer physics \cite{Velyaminov,swetnam,yzff,yzff2013}.
We choose the strategy of starting from a given seed conformation of
the knot to be investigated.
The original seed conformation is later  changed
using a set of 
random transformations called the pivot moves~\cite{pivot}.
Random transformations are used both for the  equilibration of the
seed  and for sampling a statistically relevant set of knot conformations.
The topology after each transformation is preserved with the help of
the PAEA method, first introduced in \cite{yzff}.
Our studies are performed on a simple cubic lattice and in the so-called 
stress ensemble, in which the tensile forces and their application
points are regarded as the known
”thermodynamical” parameters. The
variables conjugated to the forces
are the average distances at equilibrium
of the application points with respect to 
a set of reference points, lines or planes. These distances are
evaluated numerically.
The Hamiltonian is that already
used in Refs.~\cite{yzff,yzff2013} with the addition of a mechanical term
describing a constant force $F$ pulling the knot at one point while another
point is anchored at one site of the lattice.

The goal pursued in this work is to understand how a polymer knot 
behaves under 
stretching at different temperatures. So far, there have been few
studies of this kind.
When no  external forces are applied,
an exhaustive investigation of the thermal properties of linear open
chains  can be found in \cite{binder}. For
very short-range interactions, as those considered in that work, it
turns out that the
heat capacity of linear open chains in a bad solvent, plotted as a
function of the 
temperature,  exhibits a single sharp peak 
corresponding to a phase transition from a crystallite state to 
an expanded coil state. Such phase transitions occur at low
temperatures and are usually related to states which are very compact
and thus statistically
rare. 
The thermal properties of  unstretched polymer knots on a single
cubic lattice have been the subject of
Refs.~\cite{yzff,yzffappleb,yzff2013}. Also in the case of knots, a close
inspection of the conformations  shows that, in a
bad solvent, 
ordered compact structures are forming at low temperatures, corresponding
to the crystallite phase. At  higher temperatures,  
the knot is found in an  expanded state.
The heat capacity is characterized by a single peak,
corresponding to the phase transition from the crystallite to the expanded
state.
The behavior of the specific energy of the
knot and its gyration radius confirms this conclusion.
The crystallite--expanded state transition  does not seem to be a
lattice artifact or a pseudo-phase  
transition as those discussed in Ref.~\cite{vogel}. Indeed, the height 
of the peak grows proportionally to the number of segments $N$ composing
the knot at least up to the maximum length we could check so far ($N=2100$).
Similar results were also found in the work \cite{Velyaminov}, where
the statistical properties of unknotted rings, the so-called unknots,
have been studied.
Up to now it is still unclear
if there are other 
phase transitions
occurring starting from phases in which
the polymer knot finds itself in a particular set of
conformations that are 
extremely rare.
The existing studies of very short polymer rings with no knotting
confirm the presence of 
a single peak in the heat capacity  \cite{Velyaminov,yzff}.
The case of knotted rings is complicated by 
the difficulties of sampling extremely compact conformations, which
will be discussed in more details  in Subsection~\ref{ccc}. 

The phase diagram of polymer knots becomes more interesting  when
tensile forces 
are applied.
The heat
capacity and  other observables can then be plotted 
as a function of both the temperature $T$ and the force $F$.
As already shown in \cite{swetnam},
in this extended two-dimensional space of
parameters new phases appear.
In Ref.~\cite{swetnam}, however, a different setup
has been used, in which the variable conjugate to the force $F$ is the
height of the knot along the $z-$axis. This setup is  related to
the compression and
elongation of a knot rather than to stretching.
The two approaches are not equivalent.  In our case the
expectation 
values of observables like the specific energy, the specific heat
capacity and the gyration radius are symmetric under the change of
the orientation of the 
force $F$. 
This is not true within the approach of \cite{swetnam}, because by
inverting the sign of $F$ one passes from compression to extension,
which clearly are not equivalent processes. 
Of course, the properties of the knot under stretching should be
similar to those found in
\cite{swetnam}  when the polymer is elongated. Indeed, we have verified
that this is the case.
As a consistency check, we have also reproduced part of the results of
\cite{swetnam} using their setup.

It turns out from our computations  that the phase diagram of short
polymer knots is quite diversified and that the topology of the knot
has a strong impact on its behavior.
For small stretching forces,
the system undergoes only one phase transition
from a crystallite state to an expanded state. This
transition occurs when the temperature  increases and it was already
found in \cite{yzff}.
Additionally, there
is another phase transition
which takes place at any given temperature when the knot is pulled
by a tensile force.
This transition consists in the passage from an
unstretched phase (crystallite or expanded depending if the temperature is
low or high) to 
a stretched phase. 
At low temperatures and for high stretching forces, there is
 an additional transition from a stretched but more compact phase to a
stretched and more expanded phase.

We have analyzed also the possible effects of
topology and size on the mechanical and thermal properties of polymer
knots.
The case of  the thermal properties of unstretched polymers has
already pointed out that topology  plays 
an important 
role when polymer is short and the monomers are strongly interacting,
see e.~g. Ref.~\cite{yzff2013}. In that reference it was also
established that the behavior of a knotted
polymer ring starts to be 
independent of the knot type when its length exceeds  approximately
that of  $400$ lattice units~\cite{yzff2013}.
The knots considered in this paper are much shorter than that and,
indeed, we have found that there are effects related to topology which
are relevant and visible in the plots of the computed observables. 
Strong topological effects on the mechanical properties of
knots are related to the fact that, in more complex knots, the polymer
trajectories  need to perform more turns in order to obey the
topological constraints. For this reason, the number of possible
conformations is limited and it is much harder to stretch
complex knots than simpler ones.
Knots that can be constructed with the
same minimal number of crossings like $5_1$ and $5_2$ do not show
many differences in their behavior. We find also that the strength of 
the topological effects fades out very rapidly when the polymer length
is growing.
These facts can be seen from the plots of the specific energy, specific
heat capacity and gyration radius, whose expectation values have been
computed in the present work.


The presented material is organized as follows. The
setup of the simulations performed, including the Wang-Landau sampling
strategy and the 
formulas used to compute the 
observables, will be explained in Section~\ref{method}.
The set of observables consists in the specific energy, specific heat
capacity and 
gyration radius. The results concerning
the mechanical properties are presented in
Section~\ref{result}. In order to detect possible  effects of topology
on the behavior of polymers,
several types of knots of different lengths have been considered:
the unknot $0_1$, the trefoil knot $3_1$,
the knots $5_1$, $5_2$ and $7_1$.  The size
effects have been investigated  focusing on polymer rings of different
lengths but with the same topology, namely that of the trefoil knot
$3_1$. Finally, in 
Section~\ref{sectionVI} our conclusions are drawn and possible
generalizations of this work are 
discussed.
\section{Methodology} \label{method}
\subsection{Polymer model on a simple cubic lattice}
In this work, the mechanical properties of polymer knots under stretching
are studied on a simple cubic lattice. 
The Hamiltonian $H_X$ of a polymer knot composed by
$N$ segments
can be written as follows: 
\begin{equation}
H(X)=m\varepsilon-Fd_z\label{hx11}
\end{equation}
The symbol $X$ represents
a suitable
set  of variables needed to describe the conformation of the knot in space.
The first term in the right hand side of Eq.~(\ref{hx11}) 
describes very short-range interactions between pairs of
non-bonded monomers.
$m$ is  the number of contacts, i. e. the number of pairs of non-bonded
monomers in a given conformation $X$  whose reciprocal distance
amounts to
 one lattice unit.
$\varepsilon$ is the energy cost for one contact. If $\varepsilon>0$ 
the interactions are repulsive and correspond to the situation in
which the polymer is in a good solvent.
If $\varepsilon<0$, instead, the attractive interactions typical of
bad solvents are obtained.

\begin{figure}
\begin{center}
\includegraphics[width=0.5\textwidth]{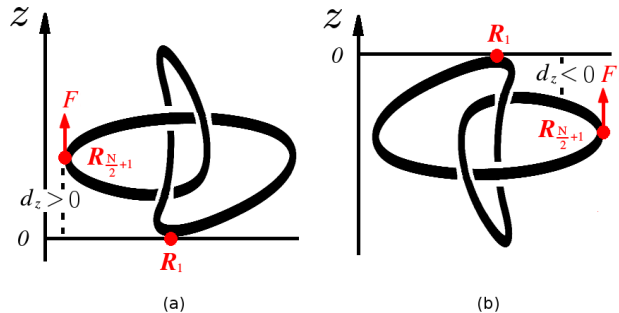}
\caption{The known force $F$ is applied to the point $\bos R_{\frac
    N2+1}$ of a trefoil knot. The orientation of the force $F$ is
  along the direction of positive $z$. (a) The value of $d_z$ is
  positive, i.~e. $d_z>0$; (b) The value of $d_z$ is negative,
  i.~e. $d_z<0$.} \label{force}  
\end{center} 
\end{figure}

The second term of Eq.~(\ref{hx11}) is the
potential associated to a constant force $F$ directed along the $z-$axis
and applied to the point $\bos R_{\frac N 2+1}$ of the knot. Here
$\bos R_1,\ldots,\bos R_N$ denote the locations of the $N$ monomers
composing the knot. The first monomer $\bos R_1$ is fixed in the
origin.
$d_z$ is the $z-$component of the vector $\bos R_{\frac N2+1}-\bos R_1$,
See Fig.~\ref{force} for a pictorial representation of our set-up.

\begin{figure}
\centering
\includegraphics[width=0.5\textwidth]{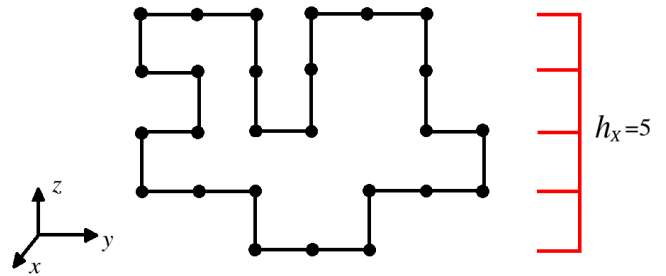}
\caption{The height $h_X$ of the conformation $X$ along the $z$
axis.} \label{height}    
\end{figure}

\begin{figure}
\centering
\includegraphics[width=0.4\textwidth]{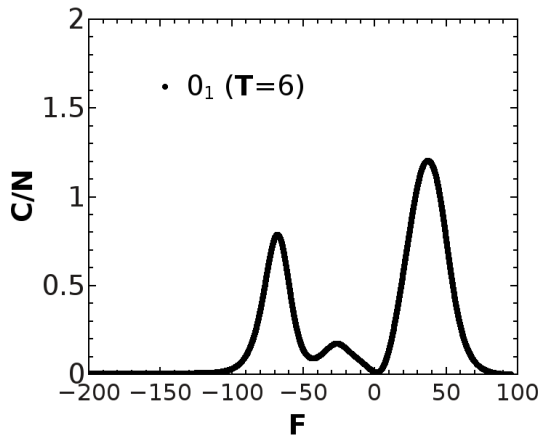}
\caption{Plot of the specific heat capacity of an unknot $0_1$ with 36
segments 
with respect to the force $F$. The force is directed along the
$z−$axis. Its orientation depends on the sign of $F$.} \label{swe-mode}   
\end{figure}

To check the consistency of our code, we repeat the calculations of
the specific 
heat capacity for an unknot presented in \cite{swetnam}.
In that work,
 the thermodynamic parameters are the force $F$ directed along the $z-$axis
and the
height $h_X$ in the $z$ direction of the conformation $X$, see
Fig.~\ref{height}.
Clearly,  $h_X>0$ while
$F$ can take both positive and negative values, corresponding to
stretching and compression respectively.
With this setting, the phase diagram of
lattice polymer rings 
under stretching has been determined in Ref. \cite{swetnam} in the
case of a few knots subjected to attractive interactions
($\varepsilon<0$). Our results are 
quantitatively coinciding with the calculation of \cite{swetnam}. For
example, the specific heat
capacity of an unknot with 36 segments displayed in
Fig.~\ref{swe-mode} is in 
agreement with the corresponding result appearing in Fig. 16 of
Ref.~\cite{swetnam}. 

\subsection{Sampling strategy}
The sampling strategy  pursued in this work consists in
starting from  a given seed conformation and then changing it with random
transformations in order to obtain a statistically relevant set of
conformations. The sampling itself is performed applying the variant
of the Wang-Landau Monte Carlo algorithm of Ref.~\cite{newwl} and that
will be discussed below. 
Before applying the Monte Carlo algorithm, the initial seed
conformation is subjected to random changes in order
to equilibrate it.
The equilibrating procedure and the criteria for deciding when equilibrium
is reached
have been illustrated in \cite{yzff}.
The random
transformations used in this work both for equilibration and sampling
are  the so-called pivot moves~\cite{pivot}.
Although a formal proof of the ergodicity of these transformations in
the case of knots does not yet exist, our previous simulations
provide enough evidence  about it \cite{yzff,yzff2013}.
It is possible to construct pivot moves that are able to change an
arbitrary number of contiguous segment $k< N$ of a polymer knot.
When the polymer conformations are swollen, higher values of
$k$  are better,
because a larger part of the knot is changed and in this way
the convergence of the Wang-Landau algorithm is sped up.
For very compact conformations, instead, small values of $k$ are preferable, 
because the likelihood that a pivot move breaks the topology of the
knot grows  with  $k$ and becomes very high when monomers are densely
packed together. If too many  pivot moves 
must be rejected to avoid that the topology is modified, the time
necessary to finish successfully a simulation  may increase considerably.
It has been found in \cite{yzff2013} that a convenient approach
consists to choose
pivot moves in which $k$ may vary randomly within a given interval of
values.

In general, pivot moves with $k>3$ are potentially violating the
topology of knot. In order to preserve it,
it will be used here the PAEA (Pivot Algorithm and
Excluded Area) method, which has been developed in \cite{yzff}. As
shown in \cite{yzff}, the time required within this method 
to avoid topology changes during the sampling procedure is
proportional to the number of segments of a given polymer. This feature
makes the PAEA method one of the fastest method on the market. 

The goal of the sampling strategy explained above is
to determine the density of
states of a polymer knot.
The density of states is the crucial ingredient
in the calculation of the averages of
any macroscopic observable in the microcanonical ensemble.  

\subsection{The sampling algorithm}\label{ccc}
The Wang-Landau method consists in a step-by-step procedure that
allows to derive within the microcanonical ensemble the density of
states $\phi_E$  in a given interval of the energy $E$. At each
step, the accuracy of the computation of $\phi_E$ depends
on a parameter called the modification factor \cite{wl}, which must
decrease in a suitable way in order to increase the degree of the accuracy. 
In the original version of the method, the modification factor starts from
the initial value $f_0=e$ and is reduced  at each further step
$\nu=1,2,\ldots,\nu_{max}$ according to the rule
$f_\nu=\sqrt{f_{\nu-1}}$ \cite{wl}.

A few variants aiming to improve the Wang-Landau algorithm first established in
\cite{wl} have 
been recently proposed.
For instance, in \cite{1twl} it
has been shown that  the
way in which the modification factor is reduced affects the
convergence of the algorithm to the exact value of $\phi_E$. An
improved prescription for
decreasing the modification factor, called  the $1/t$ Wang-Landau
method, has been presented in \cite{1twl}.
For our purposes, suitable algorithms are those which take into
account
the fact that 
 polymer
knots are systems characterized by a rough energy landscape
and have an energy range that is a priori unknown.
The reason for which the energy range is undetermined is that, even
for a simple Hamiltonian like that of Eq.~(\ref{hx11}),
the analytical calculation  of the maximum
number of contacts $m$ of a knot on a  cubic lattice is difficult.
When attempting to find numerically the highest
value of 
$m$, which in the case of attractive interactions and fixed force $F$
corresponds to the
absolute minimum of the Hamiltonian (\ref{hx11}),
one encounters several, if not many relative minima, from which it is
not easy to escape using random transformations. As a consequence,
occasionally the 
system  gets trapped at some point of relative minimum.
All that seems to suggests that a polymer
knot could have  a rough energy landscape with
a complex structure.
The complexity of the energy  landscape is related to the problem of
sampling
extremely rare conformations  mentioned in the Introduction. Indeed,
conformations with very high contact number, corresponding to 
energy minima
when the interactions are attractive, have to be considered as  events 
that are very rare and thus hard to be sampled \cite{yzff2013}.
The existence of such  events slows down  the
calculation of the density of states considerably. For instance, even
in the case of 
a short polymer with 
$N=70$ segments, a conformation with $m=84$ contacts is normally
discovered after   generating
tens of  billions of conformations. In this situation, it is easy to
realize that, without a careful
treatment of rare events, a  rare conformations may
appear for the first time when the simulation is already in an
advanced stage, i.~e. when it
is very likely that
the modification factor will be  small due to
the reduction 
procedure briefly discussed
before. 
If this happens, the algorithm is trapped into the rare configuration
until the density of state $\phi_E$ for the corrisponding energy value
is computed. This calculation must be performed with a  high accuracy
because  the modification is small and thus
requires a very long time.
A simple solution to this problem is
 to set a
  cutoff in the maximum value of $m$ \cite{yzff2013,binder}. 
In this way,
the
sampling of extremely compact or swollen conformations is
prevented. The disadvantage of this procedure is that
the exclusion of conformations with a higher number of contacts
may result in a poor description of the thermal properties
of the polymer at low  temperatures if
the interactions are attractive
\cite{binder}. Moreover, 
with an inappropriate choice of the
energy cutoffs, those phase transitions that are taking place at low
temperatures and thus involve very compact conformations, see for
example \cite{binder,wl2}, may be  neglected.

Motivated by the need of dealing with the above difficulties,
in the present work the conformations of polymer knots
will be sampled with the help of the variant of the Wang-Landau
method explained in Ref.~\cite{newwl}. This variant  has been
explicitly developed in order to cope with
 systems with a rough energy landscape and a 
energy range which is not fixed as it is in the present situation.


Finally, let us note that, according to the Hamiltonian of
Eq.~(\ref{hx11}), the energy of a conformation
$X$ depends on the force $F$,
 the number of contacts $m$ and the distance $d_z$. In the following,
it will be convenient to compute the density of state  
 as a function of 
$m$ and $d_z$. Thus, the density of state will be
denoted hereafter with the symbol $\phi_{m,d_z}$.  
 During the 
sampling procedure,
the transition probability from a microstate $X_{m,d_z}$ to another
microstate $X_{m',d_z'}$ will be given by:
\begin{equation}
P(X_{m,d_z}\longrightarrow
X_{m',d_z'})=\min\left[1,\frac{\phi_{m,d_z}}{\phi_{m',d_z'}}\right]
\label{probtrans} 
\end{equation}
The modification factor $F_{new}$ is choosen using the prescriptions of
Refs.~\cite{newwl,wl2} adapted to the present case:
\begin{equation}
F_{new}=p\sum_{i=1}^{N_{m}}\sum_{j=1}^{N_d}\left(\frac{M(m_i,d_{z,i})}
{\sum_{i=1}^{N_{m,d}}\sum_{j=1}^{N_d}
M(m_i,d_{z,j})}\right)^2-\frac{p}{N_mN_d} \label{newvaf}   
\end{equation}
In the above equation we have supposed that the possible
values of the number of contacts $m$ and of the distance $d_z$
are labeled by indices $i=1,\ldots,N_m$ and $j=1,\ldots,N_d$
respectively.
Of course, for a knot with $N$ segments, $N$ being an even integer, we
have that 
$-\frac N2< d_{z,j}< +\frac N2$.
Moreover $m_i=0,1,\ldots,m_{max}$.
For a trefoil knot with $N=70$ the maximum number of contacts found
after  generating  one hundred of billions of random conformations
is $85$.
For a general knot of arbitrary length, both the upper limit
$m_{max}$ of the number of 
contacts and the boundaries of the interval of the allowed
values of $d_z$ are difficult to be determined. 
For this reason, the boundary limits of both $m$
and $d_z$ are kept open during the whole simulation. This means that the
integers $N_m$ and $N_d$ can change whenever
it is accepted a new conformation,
in which the
number of contact, the distance $d_z$ or both 
 have not been  
detected in the previously sampled conformations.
For example, $N_m\longrightarrow N_m+1$ if the new conformation is
characterized by a value of $m$ that was not encountered before.
$M(m_i,d_{z,j})$ is the energy histogram. It depends on $m$ and $d_z$
for convenience,
but the
energy $H(X_i)$ of the 
conformation $X_i$ can be easily derived once the values of $m$
and $d_z$ are known: $H(X_i)=m_i\varepsilon-Fd_{z,i}$.
The initial value of $M(m_i,d_{z,j})$,  before the first conformation
with $m=m_i$ and $d_z=d_{z,j}$ is discovered, is set to zero.
When at some step of the simulation
a conformation $X$ is accepted with number of contacts $m=m_i$ and
distance $d_z=d_{z,j}$, then $M(m_i,d_{z,j})$ is updated as follows:
$M(m_i,d_{z,j})\rightarrow M(m_i,d_{z,j})+1$.
Simultaneously, the value of the modification factor $F_{new}$ is set
using Eq.~(\ref{newvaf}).
The prefactor $ p$ appearing in that equation is
 a tuneable parameter. Its
value depends on the system under
investigation. In our simulations,  $p$ has been set to be equal to one.
The prescription for reducing the modification factor $F_{new}$ given in
Eq.~(\ref{newvaf}) has been obtained in
\cite{wl2} by  requiring that the convergence of the Wang-Landau
algorithm to the density of state, estimated using a suitable function
of the probability  distribution of the data during the sampling, is
optimized. 
This variant of the Wang-Landau  algorithm has been proved to be particularly
effective in the case of short polymer knots and with densities of
states depending on more than one variable \cite{swetnam}.  
It is easy to show that Eq.~(\ref{newvaf}) can be cast also in the form:
\begin{equation}
F_{new}=p\sum_{i=1}^{N_m}\sum_{j=1}^{N_d}\left(\frac{M(m_i,d_{z,j})}
{\sum_{i=1}^{N_m}\sum_{j=1}^{N_d}M(m_i,d_{z,j})}-\frac{1}{N_{m}N_d}\right)^2
\label{newvaf2}   
\end{equation} 
so that $F_{new}\ge 0$ as it should be.
In the optimized Wand-Landau algorithm of \cite{newwl}, the transition
probability 
$p(i\rightarrow i')$ is the same of that of the original method,
i. e. in our case is given by Eq.~(\ref{probtrans}).
The difference is that the modification factor 
$F_{new}$ shown in Eq.~(\ref{newvaf}) needs to be updated after each
Monte Carlo move, because it depends on the values of the energy
histogram.
Moreover, if the visited state is a new one, which has never been
explored before, the values 
of $N_m$ and $N_d$, which are used
to count the number of all distinct energy states, can change as
mentioned before.
The sampling procedure is continued until a flat
energy histogram of all distinct energy states is reached. 
The flatness of the histogram is assessed within a $20\%$ of
accuracy. It is easy to check that, if the histogram is ideally flat,
$F_{new}$ vanishes identically.

\subsection{Observables}

By assuming thermodynamic units in which the Boltzmann constant is set
to be equal to one, the
partition function of the studied 
system may be written as follows: 
\begin{equation}
Z(T,F)=\sum_m\sum_{d_z} \phi_{m,d_z} e^{-(m\varepsilon-Fd_z)/T}
\end{equation}
For any given value of the temperature $T$ and the force $F$, the
density of states 
$\phi_{m,d_z}$ is derived by sampling over all microstates
$X_{m,d_z}$  with fixed number $m$ of contacts and  fixed  $d_z$ \cite{swetnam}. 
By using the knowledge of $\phi_{m,d_z}$,
the average value of any observable ${\cal O}(m,d_z)$ 
depending on $m$ and $d_z$ can be easily defined:
\begin{equation}
\langle{\cal O}(m,d_z)\rangle(T,F)=\frac{\sum_m\sum_{d_z} {\cal
O}(m,d_z)\phi_{m,d_z}e^{-(m\varepsilon-Fd_z)/T}}{Z(T,F)} 
\end{equation}
For example, the specific energy and the specific heat capacity are
respectively given by:
\begin{equation}
\frac {\langle E\rangle(T,F)}{N}=\frac 1N \dfrac{\sum_{m}\sum_{d_z}(m
\varepsilon-Fd_z)\phi_{m,d_z}  e^{-(m\varepsilon-Fd_z)/T}}{Z(T,F)}
\end{equation}
\begin{eqnarray}
\frac {C(T,F)}{N}&=&\frac 1N \left[\frac{\langle m^2\rangle-\langle
  m\rangle^2}{T^2}\varepsilon^2+F^2\left(\frac{\langle
  d_z^2\rangle-\langle
  d_z\rangle^2}{T^2}\right)\varepsilon^2\right.\nonumber\\
  &&\left.-2F\left(\frac{\langle
  md_z\rangle-\langle m\rangle\langle
  d_z\rangle}{T^2}\right)\varepsilon \right]
\end{eqnarray}
where $\langle \ldots\rangle$ denotes the operation of averaging
with respect of all possible conformations $X_{m,d_z}$ at given $T$
and $F$.   
The mean square radius of gyration $\langle R_{G}^2
\rangle(T,F)$ will be computed according to the formulas:
\begin{eqnarray} 
\langle R_{G}^2
\rangle(T,F)=\dfrac{\sum_{m}\sum_{d_z}{\langle
    R_{G}^{2}\rangle_{m,d_z}}\phi_{m,d_z}  e^{-(m\varepsilon-Fd_z)/T}}{Z(T,F)}
\end{eqnarray}
where
$R_{G}^{2}=\frac{1}{2N^2}\sum_{I,J=1}^N(\bos R_{I}-\bos R_{J})^2$ and $\langle
    R_{G}^{2}\rangle_{m,d_z}$ is the average of $R_{G}^{2}$ restricted
to states $X_{m,d_z}$ with $m$ contacts and distance $d_z$.

\section{Results} \label{result}
The topological and size
effects on the mechanical and thermal behavior of polymer knots under 
stretching are the main subject of this Section.  
To check the topological effects we have considered different types of knots
with the same number of segments $N$. The size effects  have been
investigated instead by fixing the knot type and changing its length.
As already mentioned, throughout this paper thermodynamic units are used
in which $k_B=1$, where $k_B$ is the Boltzmann constant. Moreover,
 the energy is counted in units of $\varepsilon$, so that it will be
convenient to introduce the dimensionless energy unit $\mathbf T$
defined as follows:
\begin{eqnarray}
\mathbf T=\frac {T}{|\varepsilon|}
\end{eqnarray}
Also the force $F$ will be normalized introducing the rescaled quantity:
\begin{equation}
\mathbf F=\frac F{|\varepsilon|}
\end{equation}
$\mathbf F$ has the dimension of the inverse of a length.
After this rescaling of variables, it is possible to define the
normalized Hamiltonian ${E}(X)$ such that:
\begin{equation}
H(X)=|\varepsilon|E(X)
\end{equation}
where
\begin{equation}
E(X)=m\mbox{sign}(\varepsilon)-\mathbf Fd_z\label{rescaledham}
\end{equation}
\begin{figure*}
\begin{center}
\includegraphics[width=0.50\textwidth]{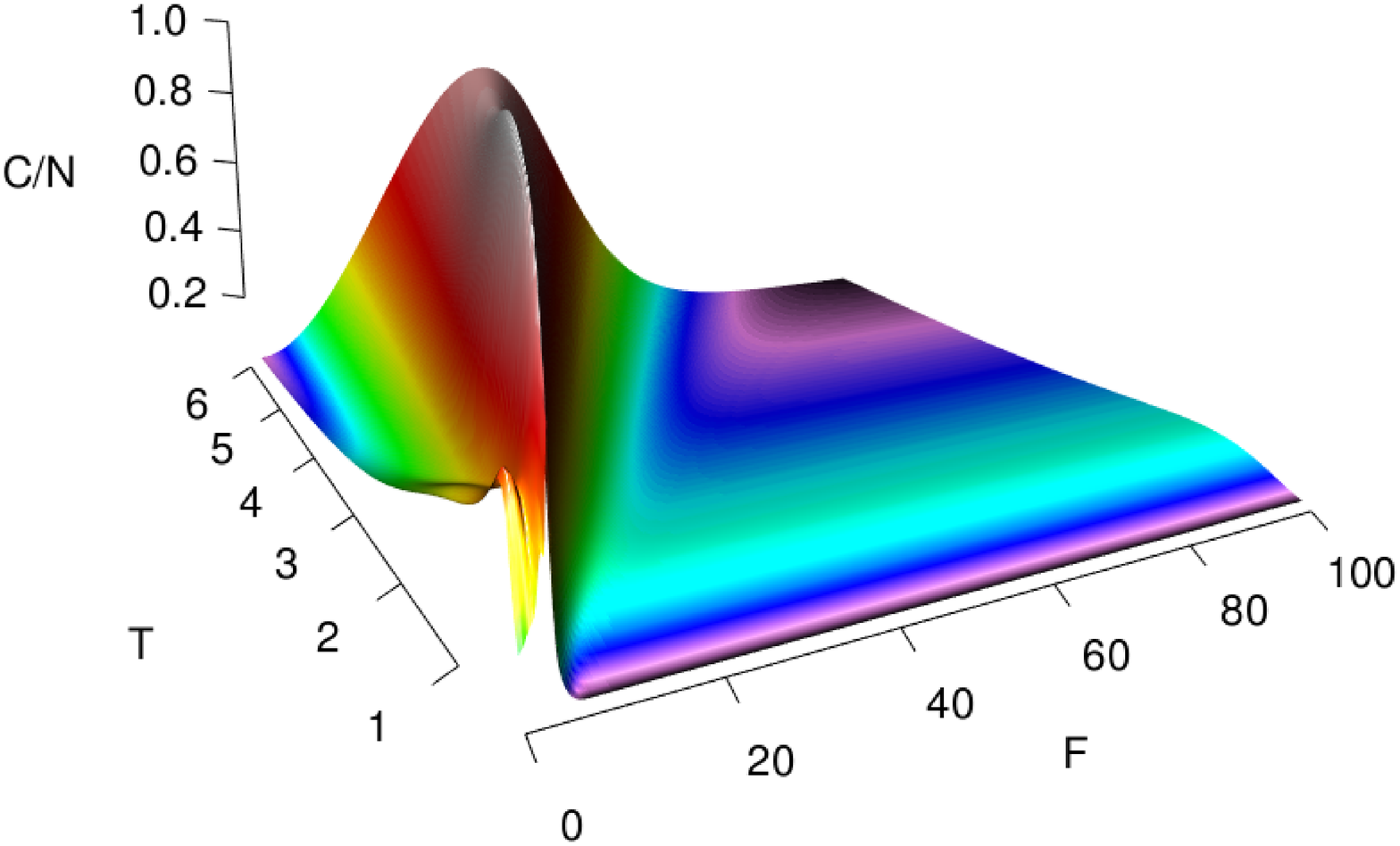}
\includegraphics[width=0.47\textwidth]{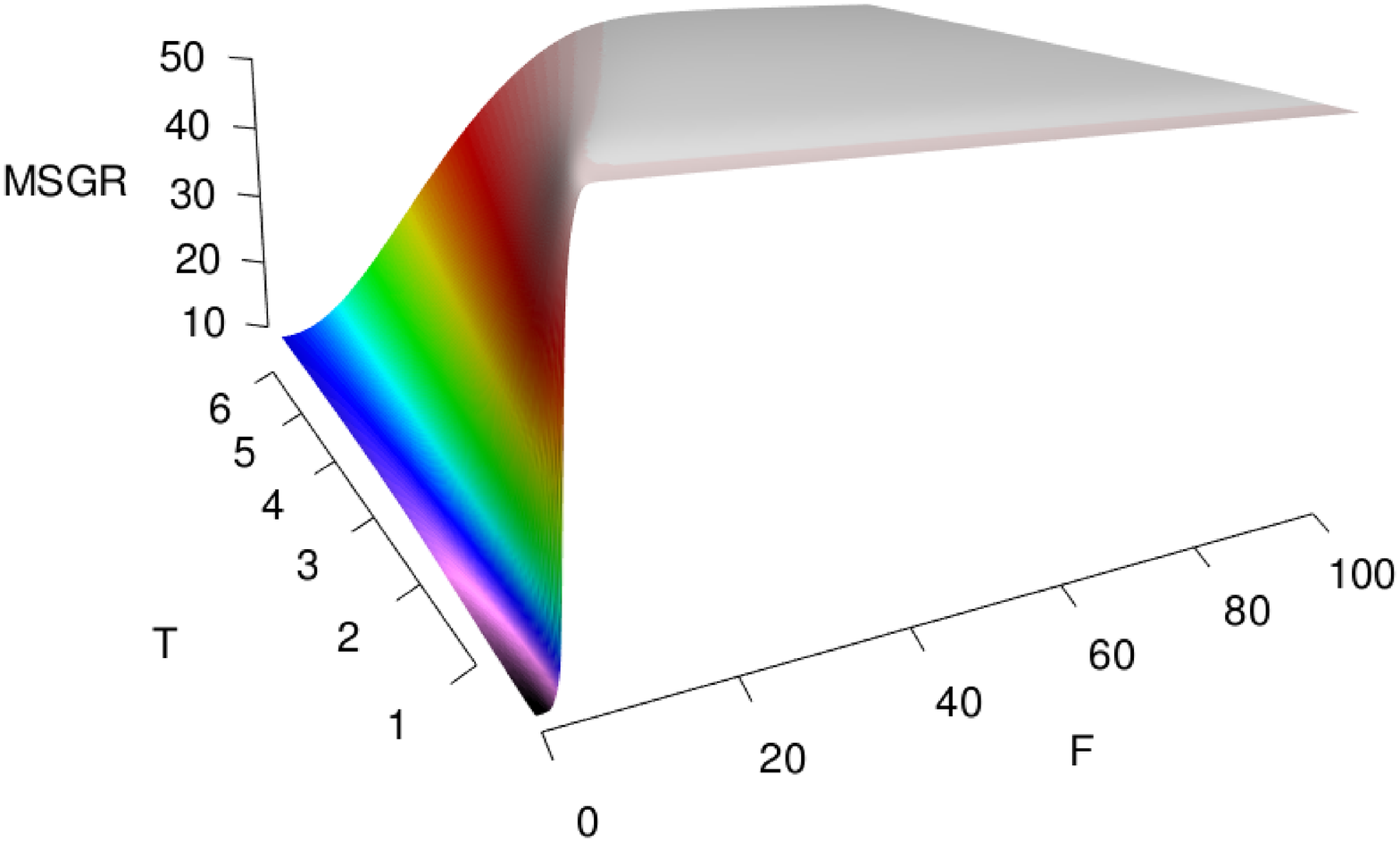}
\caption{Plot of
  (a) the specific heat capacity of a trefoil knot with $N=70$ segments
as a function of the rescaled force $\mathbf F$ and the temperature
$\mathbf T$; (b)  
  the mean square gyration radius of the trefoil. Both plots refer to the case
 of attractive forces  $\varepsilon<0$.}  \label{3t70}   
\end{center} 
\end{figure*}
Clearly we have that $\frac {H(X)}{T}=\frac{E(X)}{\mathbf T}$ and
$Z(T,F)=Z(\mathbf T,\mathbf F)$. This justifies the computation of the
observables using the rescaled units $\mathbf T,\mathbf F$.

The results concerning the topological effects will be presented in
Subsection~\ref{topo3.1}, while the effects related to size will be
the subject of Subsection~\ref{size3.2}.
We restrict ourselves to the case $\mathbf F\ge 0$, because, due to the
already mentioned symmetry of the Hamiltonian under the transformation
$\mathbf F,d_z\longrightarrow -\mathbf F,-d_z$, the plots of the studied observables
when $\mathbf F$ is negative are symmetric to those with $\mathbf F>0$
and do not provide any new information. 
\subsection{Topological effects of polymer knots} \label{topo3.1}
Three dimensional pictures are very helpful to
illustrate the general features of the behavior of the analyzed knots.
As a consequence, to start with,
  the three dimensional pictures of the 
specific heat capacity  
and gyration radius of a trefoil ($3_1$) knot 
with $70$ monomers will be discussed.
The plot of these observables
 are
displayed  in Figs.~\ref{3t70}~(a) and (b) in the range of parameters
$0\leq\mathbf  F\leq 100$ and $0.35\leq\mathbf T\leq 6$.
Attractive interactions have been assumed, i. e. $\varepsilon<0$.

\begin{figure}
\begin{center}
\includegraphics[width=0.47\textwidth]{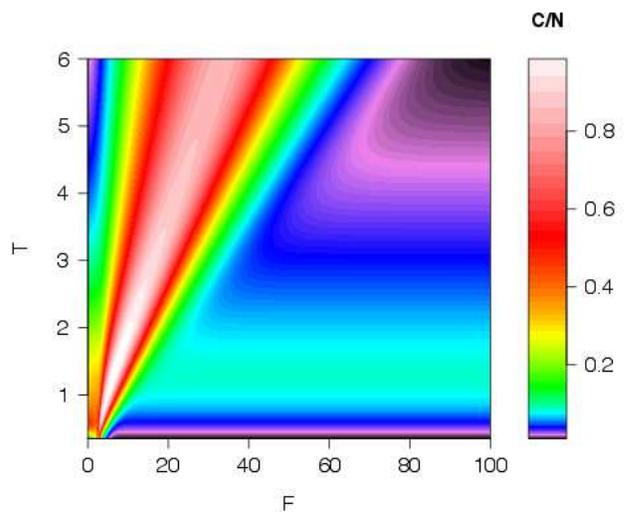}
\caption{Color shaded picture of the
heat capacity of a trefoil knot with $N=70$ segments in the $\mathbf
T,\mathbf F$ plane. In this figure it is possible to recognize a marked peak,
corresponding to the transition from a compact to a stretched
state. The peak becomes wider with increasing temperatures.
Moreover, a secondary, smaller peak is shown (cyan area extending
from $0.5\le \mathbf T\le 2$.}  \label{newfig1}   
\end{center} 
\end{figure}

From Fig.~\ref{3t70}~(a) it turns out that
the trefoil polymer knot can be found in at least three
 different phases, 
which we identify as crystallite, expanded and stretched phases.
At low temperatures and for weak stretching forces $F$, the knot is in
the crystallite state. Its conformation is compact and ordered. With
the rising of the temperature, this state 
undergoes a transition to the expanded state, characterized by a
swollen and disordered conformation. This transition
corresponds to the small peaks appearing approximately in the region
in Fig.~\ref{3t70}~(a) centered around
 $\mathbf T=1$ and such that $\mathbf F\sim 0$.
When the strength of the tensile force grows, there is a 
transition from the crystallite state (at low temperatures) or from
the expanded state (at high temperatures) to the stretched state. The
broad peak characterizing the 
the heat capacity of the trefoil knot  through the whole range of
temperatures is caused by this transition.
This conclusion is also  confirmed by the 
results of the calculation of the  mean square gyration radius
displayed in
Fig.~\ref{3t70}~(b), which clearly shows that, for fixed $\mathbf T$
and $\mathbf F$,
the growth rate of the gyration radius attains its maximum exactly in
the position of the  $\mathbf T,\mathbf F$ plane in which the broad
peak of the heat 
capacity is occurring.
For
example, when $\mathbf T=6$, the 
peak in the heat capacity is located at $\mathbf F=33.90$, see
Fig.~\ref{3t70}~(a). Looking at Fig.~\ref{3t70}~(b), it is easy to
realize that the coordinates of the maximum of
the growth rate of the
gyration radius are the same, i. e.  $\mathbf T=6$ and  $\mathbf F=33.90$. 
From Figs.~\ref{3t70}~(a) and \ref{newfig1}
it can also be observed 
that, as $\mathbf T$ 
decreases, the peak of the specific heat capacity becomes narrower and
moves to lower values of $\mathbf F$.
Correspondingly, 
the growth of the
gyration radius near the transition is stronger at
lower temperatures $\mathbf T$ and the point of maximum growth rate
translates together with the peak of the heat capacity, see
Fig.~\ref{3t70}~(b).  
The narrowing of the peak is probably related to the
fact that the stretching process is much more violent when
the knot is in a compact state at low temperatures than at high
temperatures, when the  knot conformation is already swollen.
On the other side, 
in order to stretch a polymer, the force $F$ needs to counteract the
molecular motions of its monomers. 
With decreasing temperatures, the latter
become weaker. This explains why the position of the
peak in the heat capacity 
in the $\mathbf T,\mathbf F$ plane moves along a  curce
such that $\left.\frac{d \mathbf T_{peak}(\mathbf F)}{d\mathbf
F}\right|_{F_{peak}}>0$ as 
shown clearly in 
Fig.~\ref{newfig1}. Of course, when the stretching 
force is so high that the polymer is already almost fully stretched,
the stretching process gives negligible effects and
the heat capacity, as well as the gyration
radius, stop to change significantly.

Besides the two transitions described before, there is also a third
and minor one with a peak in the heat capacity
centered approximately 
along the line $\mathbf T=1$ in Fig.~\ref{3t70}~(a). Its presence is
clearly pointed out
in the color shaded picture of Fig.~\ref{newfig1}. The existence of
this peak  is in agreement with the calculations of \cite{swetnam} in
the case of a figure-eight ($4_1$) knot. As a matter of fact, while
the setups are different, we expect agreement of our results with those
obtained in \cite{swetnam} when the polymer is elongated.
The minor peak persists even when the stretching force is very large.
It corresponds to a transition occurring
with the rise of the temperature when the knot passes from a
stretched and compact state to a stretched and slightly more swollen
state. 
In \cite{swetnam} there are additional phase transitions taking place
when the
values of the force $F$ are negative. These are however related to
compression, 
a phenomenon that is not reproducible by stretching and thus is not
relevant in this work.

Let's now go back to the topological effects. 
Figs.~\ref{3t70} and \ref{newfig1} suggest that, in a stretched
polymer knot subjected to short range attractive forces,  there are only two
relevant phase transitions. The minor one, which is  responsible for
the smaller peak in Figs.~\ref{3t70} and \ref{newfig1}, is
probably a pseudo-phase transition and will be neglected in the
following. 
The phase transition leading to the main peak of the heat capacity in
Fig.~\ref{3t70}~(a) can 
be observed in all studied knots  at any
temperature although, as we have seen, the maximum of the peak, as
well as the maximum of the growth of the gyration radius,
occur for different values of the stretching force $\mathbf F$,
i.~e. $\mathbf 
T_{peak}=\mathbf T_{peak}(\mathbf F)$.
As a consequence, we concentrate here on
a given fixed temperature, for instance 
$\mathbf T=6$. The 
reason of this choice is the comparison
with the results of Ref.~\cite{swetnam}. In
that work, in fact, one of the two temperatures mainly used in the discussion
was
$\mathbf T=6$.  
\begin{figure*}
\begin{center}
\includegraphics[width=0.47\textwidth]{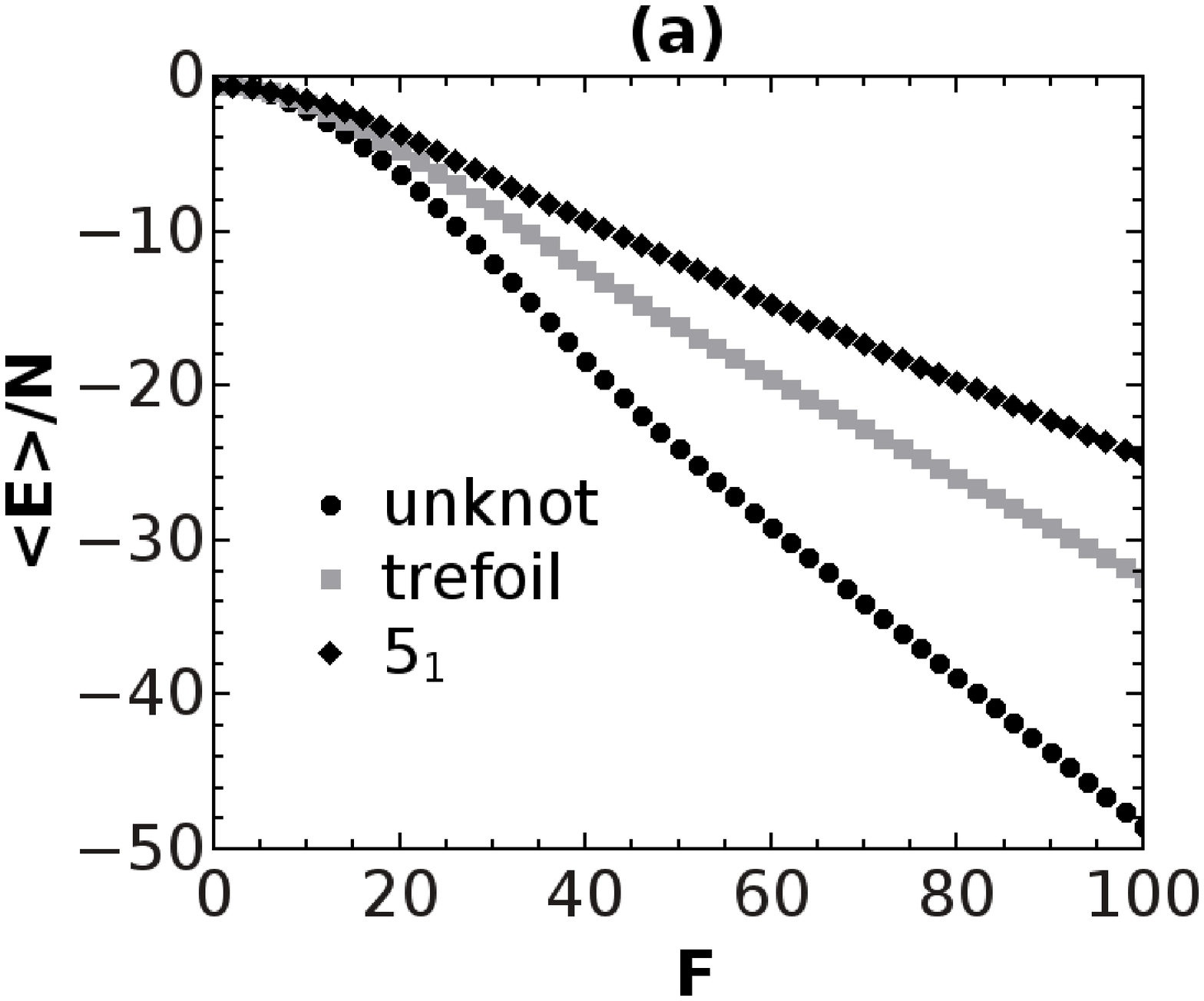}
\includegraphics[width=0.47\textwidth]{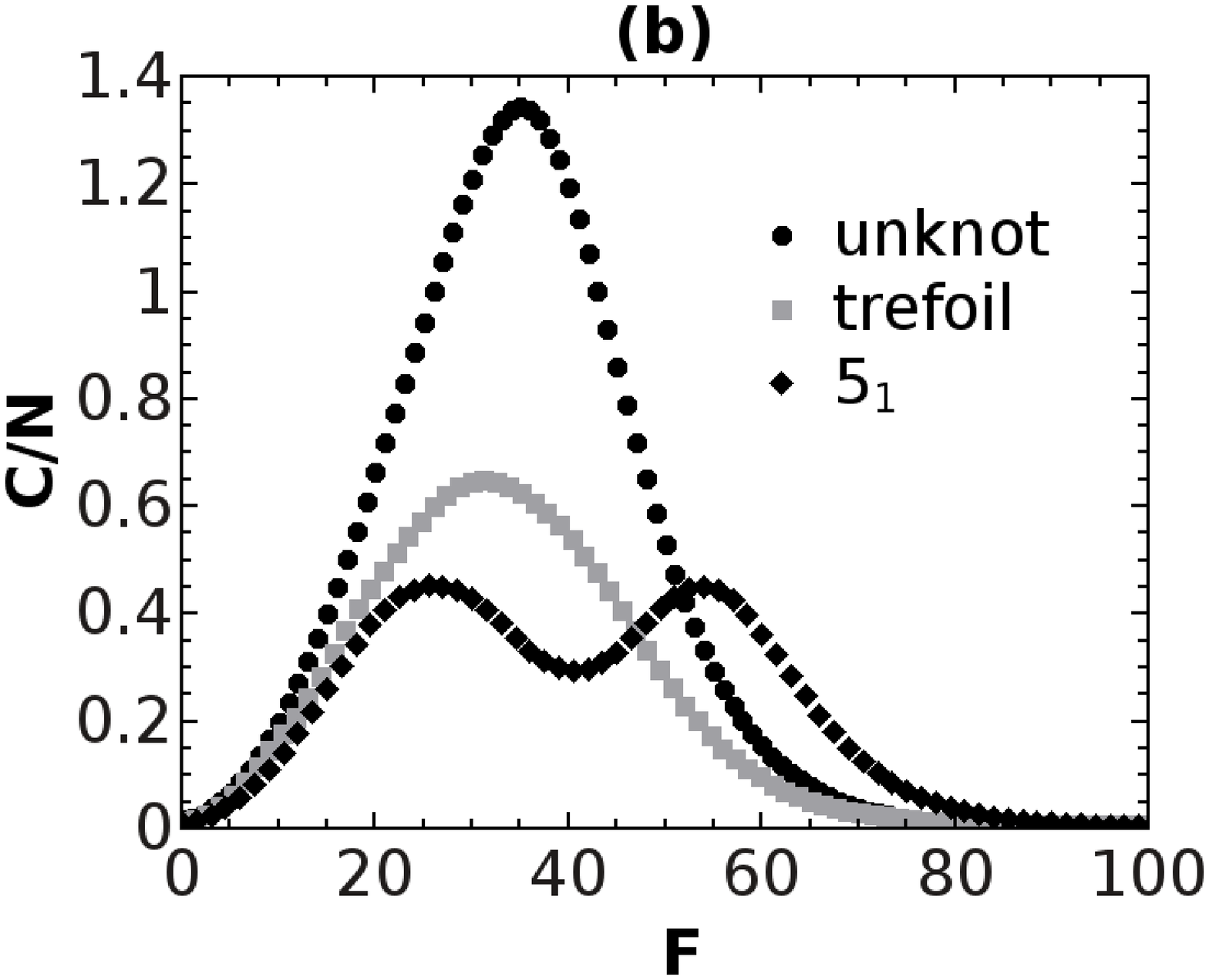}
\caption{(a) The specific energy and (b) the heat capacity of the unknot 
$0_1$ (circles), the trefoil knot $3_1$ (rectangles) and the knot $5_1$ 
(diamonds). All knots have  length of 50 lattice units. The normalized
temperature is  
$\mathbf T=6$ and 
  $\varepsilon<0$.}  \label{topology50}   
\end{center} 
\end{figure*}
To study possible topological effects, 
in Fig.~\ref{topology50} the
specific energy and heat capacity
have been plotted in the case 
of the unknot ($0_1$), the trefoil and the
knot $5_1$. All these knots have the same length $N=50$
and their monomers are subjected to attractive interactions ($\varepsilon<0$).
The specific energy, see Fig.~\ref{topology50}~(a), decreases
with the growing of the stretching force.
This is an expected result, as it is possible to realize looking at
the rescaled 
Hamiltonian ${E}(X)$ of Eq.~(\ref{rescaledham}).
Indeed, remembering that
$\mbox{sign}(\varepsilon)=-1$ if the interactions are 
attractive, we have that both terms $m\mbox{sign}(\varepsilon)$ and
$-Fd_z$ in the Hamiltonian are negative, while their absolute values
grow when the 
strength of the stretching 
force is increasing.
There are in fact two combined effects that contribute to lower the specific
energy of the system. First of all, by stretching a swollen
conformation, the number of contacts grows until a saturation value
$m_{stretch}$ is 
reached.
For example, $m_{stretch}=12$ for an unknot with $N=50$ segments.
Secondly, under increasing tensile forces, the length $d_z$ converges
in our setup to the maximum  
possible length $d_{z,max}$ of the knot. In the case of an unknot with
$N=50$, it 
is easy to see that this length is $d_{z,max}=24$.
All this explains the decrease of the specific energy
mentioned before.
For instance, when $\mathbf F=100$, we may suppose that the polymer is
almost completely stretched, so that for an unknot the average value of the
specific energy $\frac{\langle E\rangle}N\sim
-\frac{12}{50}-\frac{100\cdot 24}{50}\sim -48.22$.
Within a very good approximation, this is exactly the value of $\frac{\langle
E\rangle}N$ for $\mathbf F=100$ obtained from our calculations, see
in Fig.~\ref{topology50}~(a) the plot of the specific energy for the
unknot  $0_1$.

The data concerning the specific heat capacity are displayed in
Fig.~\ref{topology50}~(b). It is possible to realize that
the specific heat capacities of the unknot and the trefoil present
respectively a
single peak.  
\begin{figure}
\begin{center}
\includegraphics[width=3in]{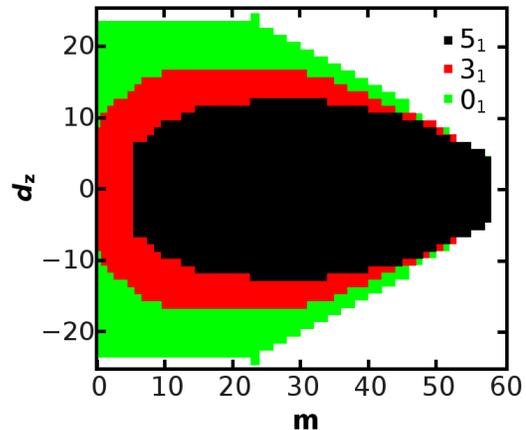}
\caption{Accessible combinations ($m,d_z$) of the unknot $0_1$ (green
  rectangles), the trefoil $3_1$ (red rectangles) and the knot $5_1$
  (black rectangles). All the knots consists of $50$ segments.} \label{m-dz}  
\end{center} 
\end{figure} 
As we have already discussed, this peak  can be interpreted as the
result of the  transition  from an expanded  phase to a stretched
phase. The case of the knot $5_1$ is exceptional
because two 
equivalent peaks are observed. The presence of this additional peak is
almost certainly a consequence of the fact that
in a knot $5_1$ with only $50$ segments
the monomers are very close to each
other and are thus strongly interacting. In this situation,
the topological effects are
enhanced.
To convince ourselves about that, it is sufficient to remember
that the minimum number of segments 
necessary to construct a knot $5_1$ on a simple cubic lattice is $34$.
In comparison, the simpler knot $3_1$ and the unknot
$0_1$ have a minimum length of  $4$ and $24$ lattice basic units
respectively 
\cite{mini}. 
It is thus expected that in such a complex and short knot
like a $5_1$ with $N=50$ segments
the expansion
process is complicated by  the topological constraints.
This fact is also confirmed by
Fig.~\ref{m-dz}, which clearly shows that
the number of possible energy states of a $5_1$
knot,
determined by the allowed values of the parameters $m$ and $d_z$ on
the lattice, is definitely less than the analogous number
in the case of a  trefoil or an unknot of the same length.
In conclusion, the appearance of the double peaks of $5_1$ is 
a topological effect  appearing due to the limitations of the knot $5_1$
in accessing  new energy states during its expansion process.
As a counter-check, 
 we have computed the specific heat capacity of a longer knot $5_1$
with  $N=70$. The result is shown in Fig.~\ref{top70} (b). As
it is possible to see, now only a single peak is observed.

By increasing the number of segments,
more complicated knots with higher minimal crossing numbers can be
formed.
In
Fig.~\ref{top70} we present for instance the plots of the specific energy and
heat capacity for  the knots $0_1,3_1, 4_1,5_1,5_2$ and
$7_1$ with $N=70$. 
\begin{figure*}
\begin{center}
\includegraphics[width=0.47\textwidth]{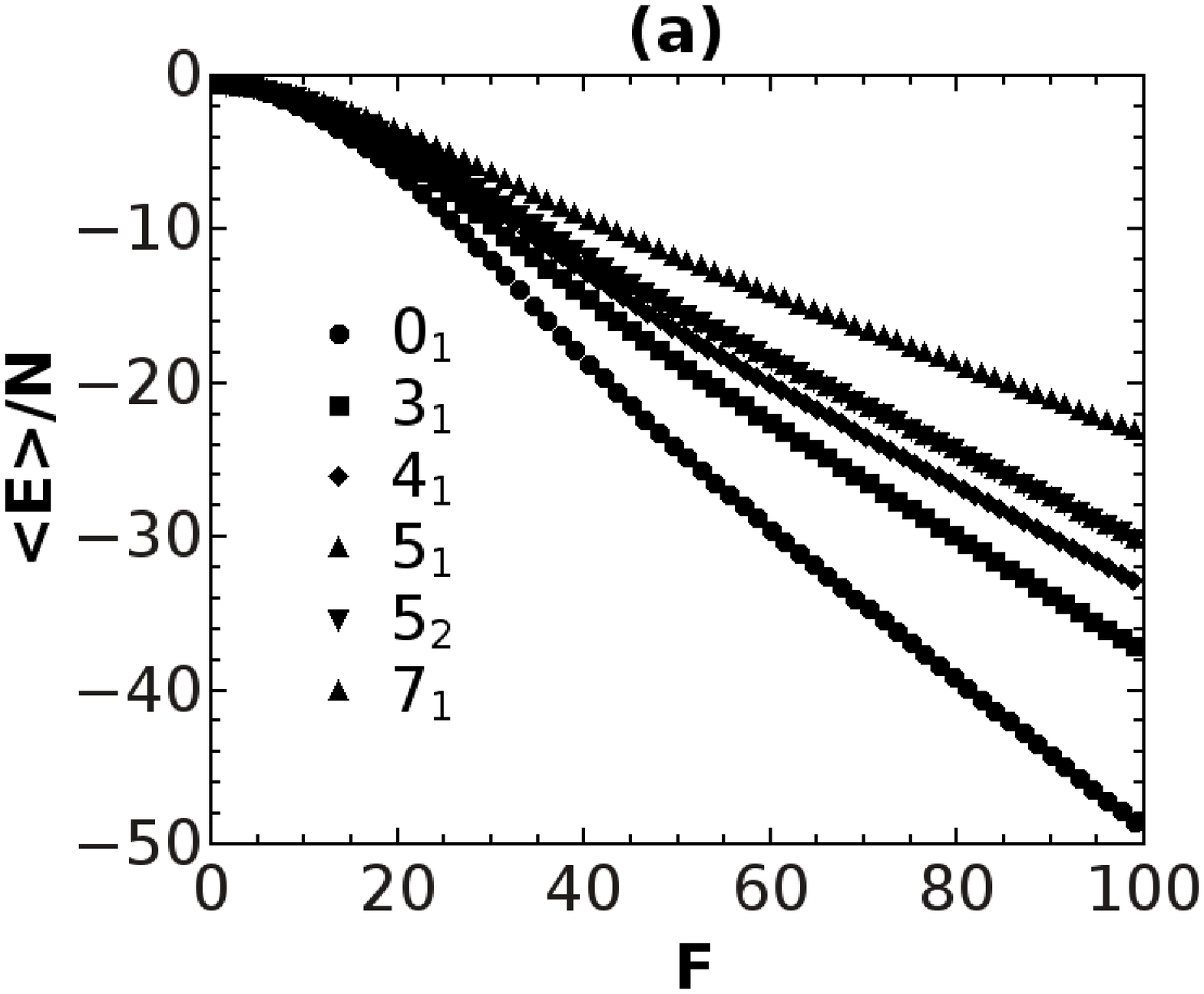}
\includegraphics[width=0.47\textwidth]{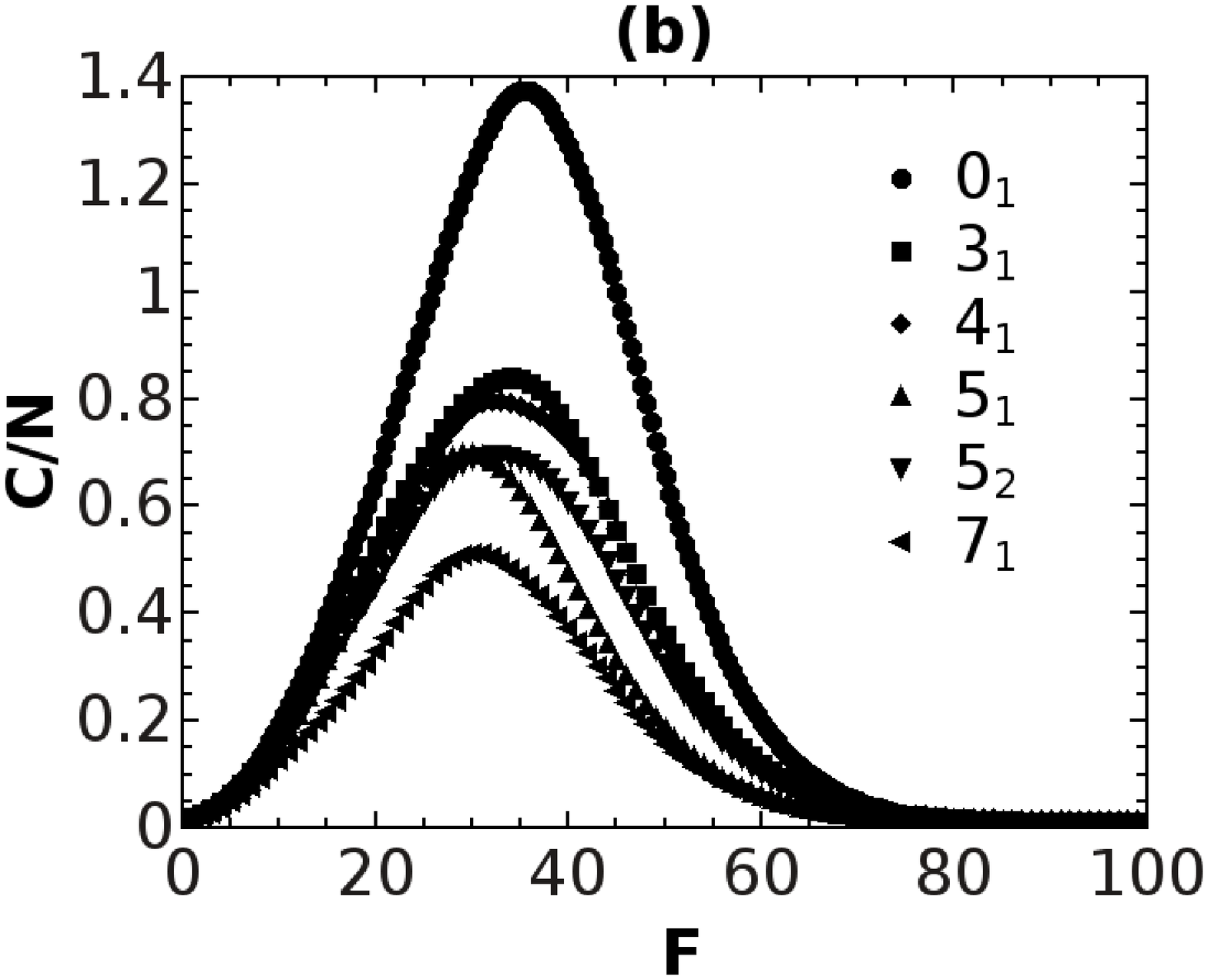}
\caption{(a) The specific energy and (b) the heat capacity of the unknot 
$0_1$ (circles), the trefoil knot $3_1$ (rectangles), the figure-eight
$4_1$ (diamonds), the knot $5_1$  
(up triangle), the knot $5_2$ 
(down triangle) and the knot $7_1$ (left triangle). All knots have 70
segments. The normalized temperature is  
$\mathbf T=6$ and 
  $\varepsilon<0$.}  \label{top70}   
\end{center} 
\end{figure*}
It is clear that topology affects the thermal properties of these polymer
knots. As shown in  
Figs.~\ref{topology50} and \ref{top70}, in fact, in the attractive case
the increasing of the knot complexity results in a 
decrease of the specific energy and the heat capacity. The reason is
that more complex knots 
tend to have more compact conformations, which have in the average a higher
number
of contacts. 
The presence of more
contacts lowers the average energy of each monomer if the interactions
are attractive, so that in turn the specific energy decreases. 
The fact that with increasing topological complexity
knots have more compact conformations has been shown in the absence of
stretching in
Ref.~\cite{yzff2013}. This conclusion is also
 confirmed when stretching is applied by the 
data on the  mean square gyration radii of the considered
 knots, see Fig.~\ref{gy70}.
\begin{figure}
\begin{center}
\includegraphics[width=0.47\textwidth]{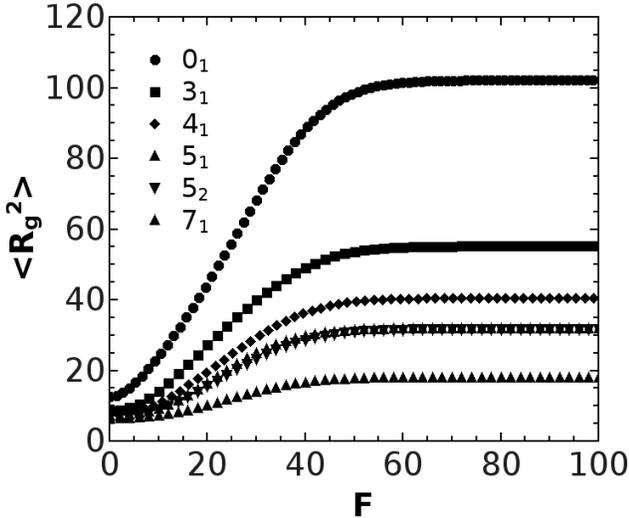}
\caption{The mean square gyration radius of the unknot 
$0_1$ (circles), the trefoil knot $3_1$ (rectangles), the figure-eight
$4_1$ (diamonds), the knot $5_1$  
(up triangle), the knot $5_2$ 
(down triangle) and the knot $7_1$ (left triangle). All knots have 70
segments. The normalized temperature is  
$\mathbf T=6$ and 
  $\varepsilon<0$.}  \label{gy70}   
\end{center} 
\end{figure} 
In particular, the gyration radii $\langle R_{G}^2\rangle(T,F)$
satisfy the inequality 
\begin{equation}
\langle R_{G}^2\rangle_{0_1}>\langle R_{G}^2\rangle_{3_1}>\langle
R_{G}^2
\rangle_{4_1}>\langle R_{G}^2\rangle_{5_1}=\langle
R_{G}^2\rangle_{5_2}>\langle R_{G}^2\rangle_{7_1}
\end{equation}
independently if the interactions are attractive -- case shown in
Fig~\ref{gy70} -- or repulsive.
Let us notice that the knots $5_1$ and $5_2$ have more or less the same
average gyration radius. This  
suggests that knots with the same minimum crossing number 
have a similar behavior.

Looking  at the plots of the specific energy and heat capacity at
fixed temperature, in particular at Figs.~\ref{sizet} and
\ref{sizet-r}, where the insets show in details the behavior when
$\mathbf F$
is small,
three regimes  
may be
distinguished in the range of forces $0\le \mathbf F\le 100$. In these regimes
the thermal properties of a
polymer knot are profoundly different:
\begin{enumerate}
\item When $\mathbf F<<\mathbf F_{peak}$, the studied system is below the
  threshold in which the transition to the stretched phase is
occurring.
In this region the decreasing of the specific energy as the stretching
force increases is relatively slow.
Moreover, the heat capacity is slowly increasing with increasing
values of $\mathbf F$.
\item Near $\mathbf F=\mathbf F_{peak}$
the system undergoes the transition
from the expanded state to the stretched state. With respect to the
previous regime, 
the
decrease of
the
specific energy  with increasing values of $F$ is more marked.
The 
peak with the maximum value of the heat capacity
is located in this region.
\item After the phase transition has been completed, the
  system finds itself in 
the stretched phase, which is characterized by large values
of the tensile force $F$. As a consequence, the mechanical term $-Fd_z$
becomes dominant in the Hamiltonian of Eq.~(\ref{hx11}) 
and the specific energy of the system
   linearly decreases proportionally to $\mathbf F$. The heat capacity goes instead to
zero because when the knot conformation is already almost stretched,
any further stretching  becomes difficult .   
\end{enumerate}
As it is possible to notice from Fig.~\ref{top70}~(b), with increasing
knot  complexity 
the height of the peak in the heat capacity is decreasing and its
width is getting narrower. This
is because in more complex knots
the stretching process is hindered by
the topological constraints, which require more turns in the polymer
trajectory than in simpler knots. As a result, the more complex the
knot, the closer to each other stay the
monomers.


\subsection{Size effects of polymer knots} \label{size3.2}
\begin{figure*}
\begin{center}
\includegraphics[width=0.47\textwidth]{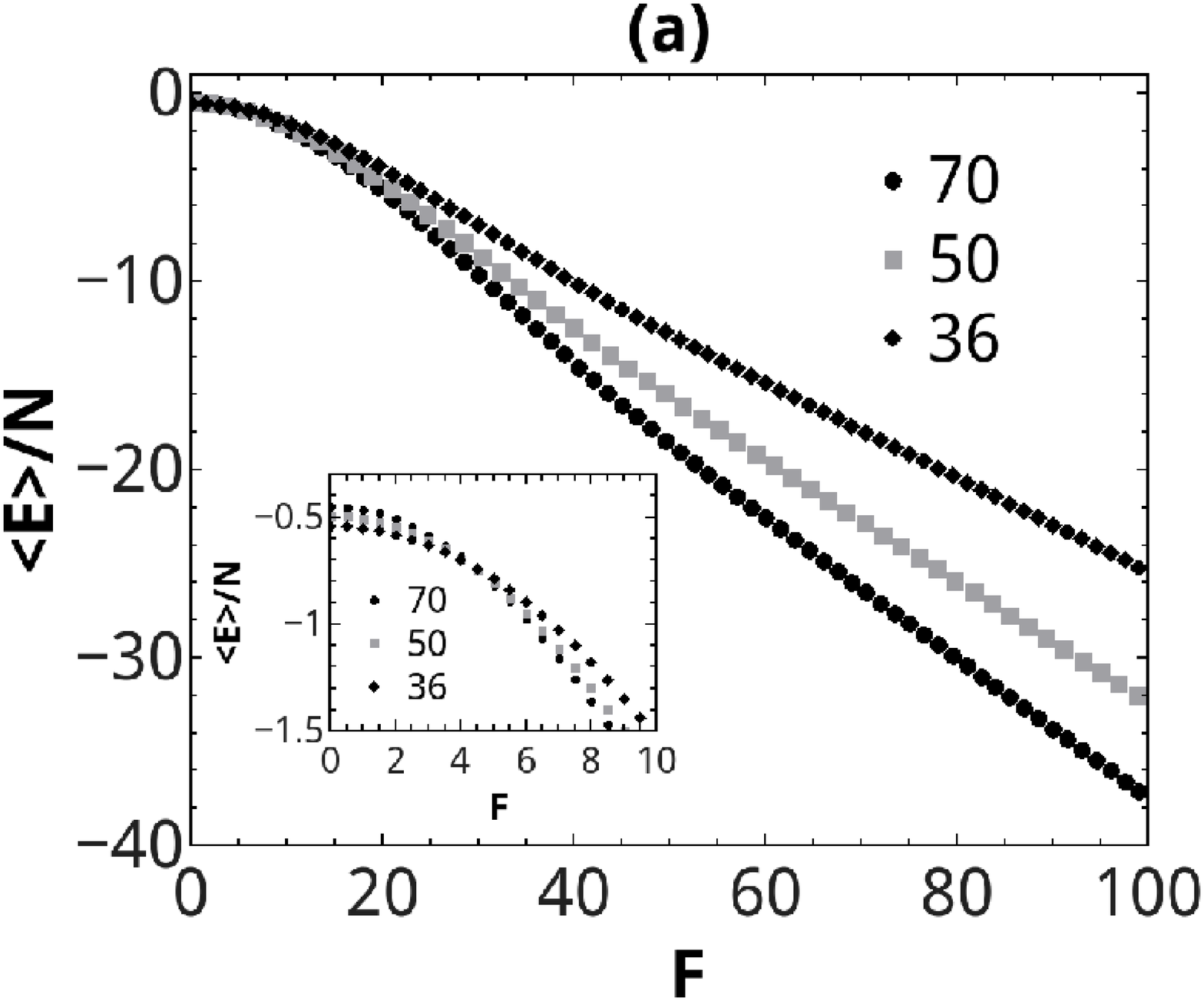}
\includegraphics[width=0.47\textwidth]{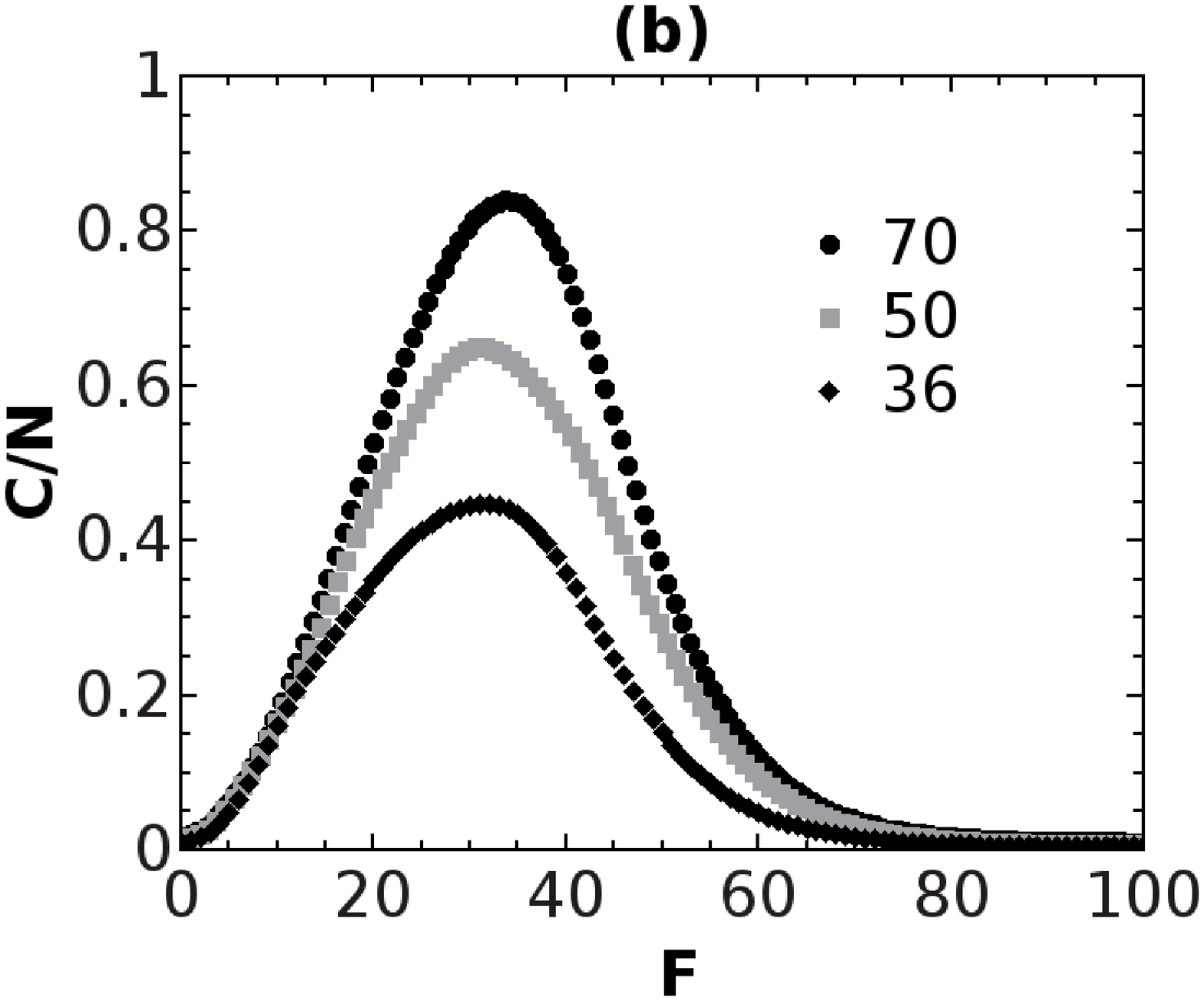}
\caption{(a) The specific energy and (b) the heat capacity of
trefoil knots $3_1$ with  lengths $N=36$ (diamonds), $N=50$
(rectangles) and $N=70$ (circles). The normalized temperature is  
$\mathbf T=6$ and 
  $\varepsilon<0$. The inset in (a) shows the details of the
behavior of the specific energy at weak tensile forces.}  \label{sizet}      
\end{center} 
\end{figure*}

To study the size effects of polymers,   we consider in this Subsection
polymers of the same knot type (the trefoil knot $3_1$) but 
different lengths. Fig.~\ref{sizet} shows the results of the
calculations of the specific
energy and heat capacity of a $3_1$ knot in the case of attractive
monomer-monomer interactions at $\mathbf T=6$. The most striking effect
which can be related to the length of the knot can be
found in the linear decay of the specific energy in Fig.~\ref{sizet}.
In the  third regime discussed in the previous
Subsection, this decay is
proportional to $F$ and, remarkably, it  becomes steeper as the knot
length $N$ increases. The reason for which the specific energy of
longer knots should be less than that of shorter polymers as observed
in Fig.~\ref{sizet}
is not entirely
intuitive, but can be explained as follows.
In the stretched phase the mechanical term $-\mathbf
Fd_z=-\frac{Fd_z}{|\varepsilon|}$ becomes 
dominant in the rescaled Hamiltonian $E(X)$ of Eq.~(\ref{rescaledham}).
Obviously, if the rescaled force
 $\mathbf F$ is fixed, the decrease
of the specific  energy can only depend on the value of $\frac {d_z}N$ and
on the topology of the considered knot. Of course, 
the maximum value of $\frac{d_z}N$ is obtained when the topology is
that of the unknot.
Indeed, if a trefoil is stretched with the same force $\mathbf F$ as an
unknot of the same length $N$, then the obtained average value of the
ratio $d_z^{3_1}(N)/N$ will be 
lower than 
that of an unknot, i. e.:
$\frac{d_z^{3_1}(N)}N<\frac{d_z^{0_1}(N)}N$. This is due to the 
fact that  in the trajectory of the
trefoil knot a part of the trajectory cannot be stretched,
because it contains the turns that are necessary in order to
satisfy the topology 
requirements. 
At this point we recall that the effects of
topology are supposed to eventually fade out
when the length of a knot becomes infinite. In this limit,  the portion of
the trajectory that is bending and twisting due to the topological
constraints  becomes irrelevant in comparison with the rest of the
knot
and the difference between
the behavior of an arbitrary knot and the unknot disappears. In other words,
the
topological effects fade out
when the number of segments $N$ is increasing. In our case, this
implies that 
the ratio $d_z^{3_1}/N$ should grow with increasing
values of $N$ in such a way that, eventually, we have that
$\lim_{N\to\infty}\frac{d_z^{3_1}(N)}N=\frac{d_z^{0_1}(N)}N$.
This growth of the ratio $\frac{d_z^{3_1}(N)}N$ with $N$ is exactly what is
observed in Fig.~\ref{sizet} and explains
 the faster decrease of the specific energy, because it is the reason
for which the dominating 
term $-\mathbf F\frac{d_z(N)}N$ in the Hamiltonian decreases with $N$.

\begin{figure*}
\begin{center}
\includegraphics[width=0.47\textwidth]{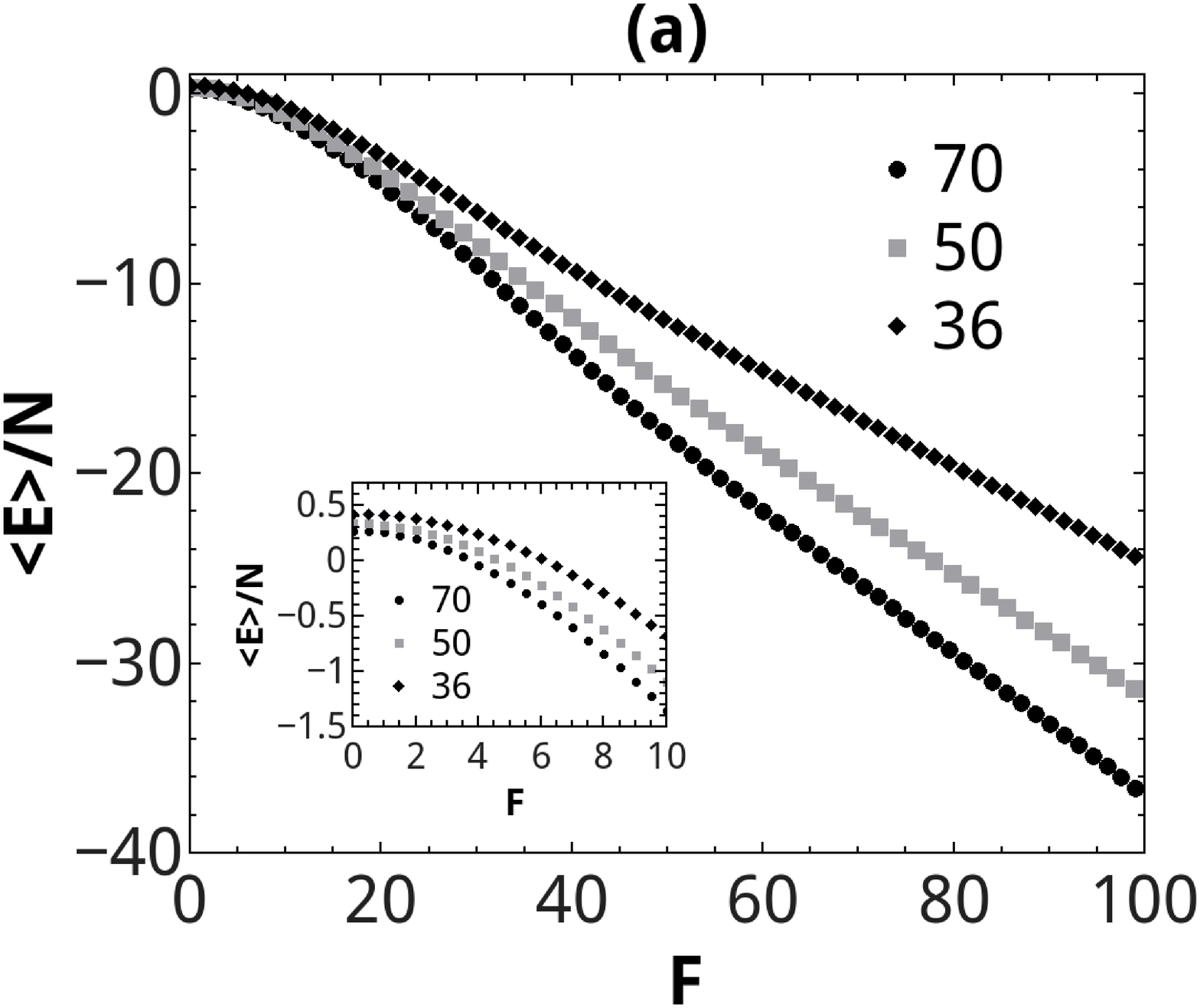}
\includegraphics[width=0.47\textwidth]{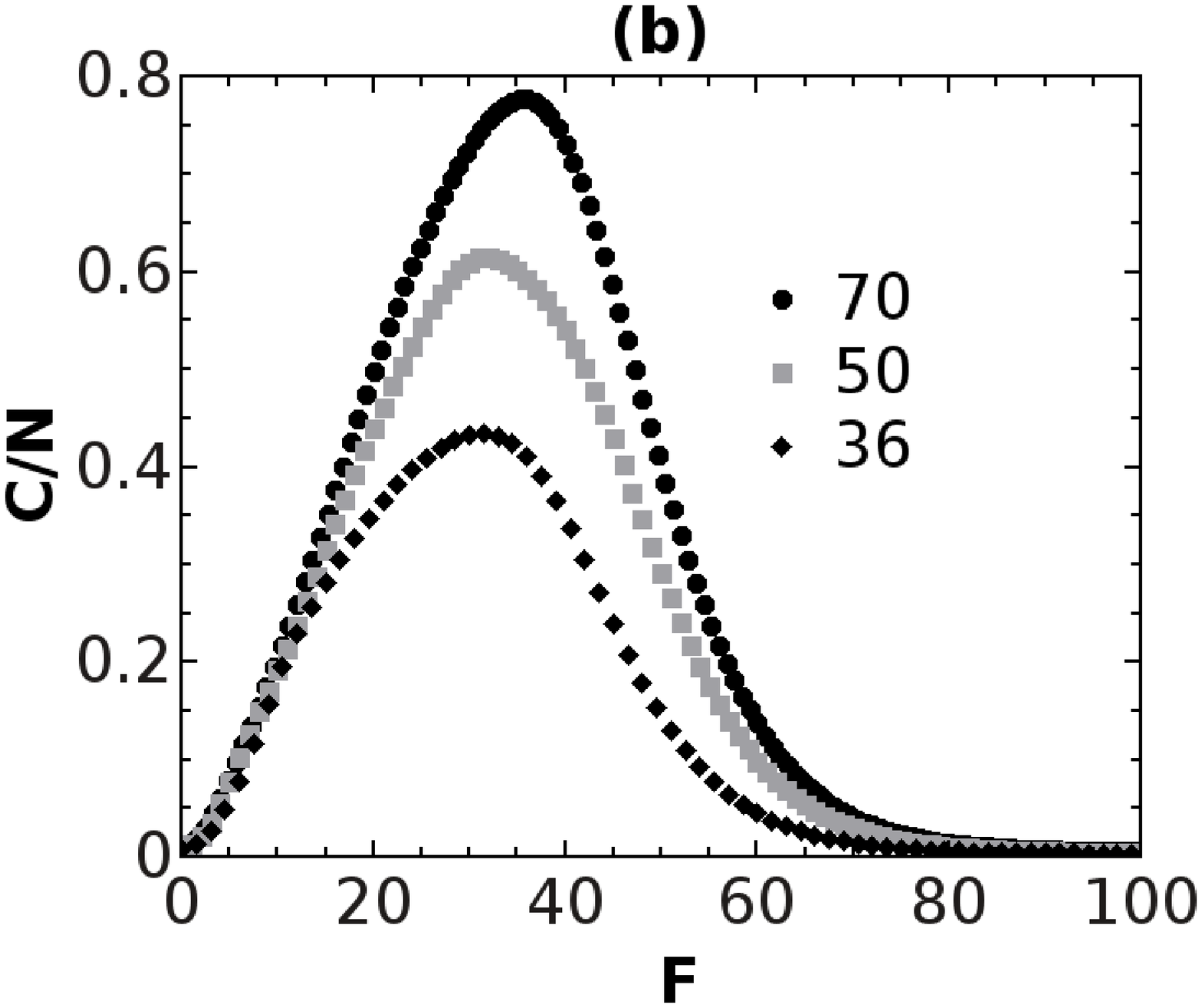}
\caption{(a) The specific energy and (b) the heat capacity of 
a trefoil knot $3_1$ considered at different  lengths $N=36$ (diamonds), $N=50$
(rectangles) and $N=70$ (circles). The normalized temperature is  
$\mathbf T=6$ and the interactions are repulsive
($\varepsilon>0$).
 The inset in (a) shows the details of the
behavior of the specific energy at weak tensile forces.
}  \label{sizet-r}   
\end{center} 
\end{figure*}

\begin{figure*}
\begin{center}
\includegraphics[width=0.47\textwidth]{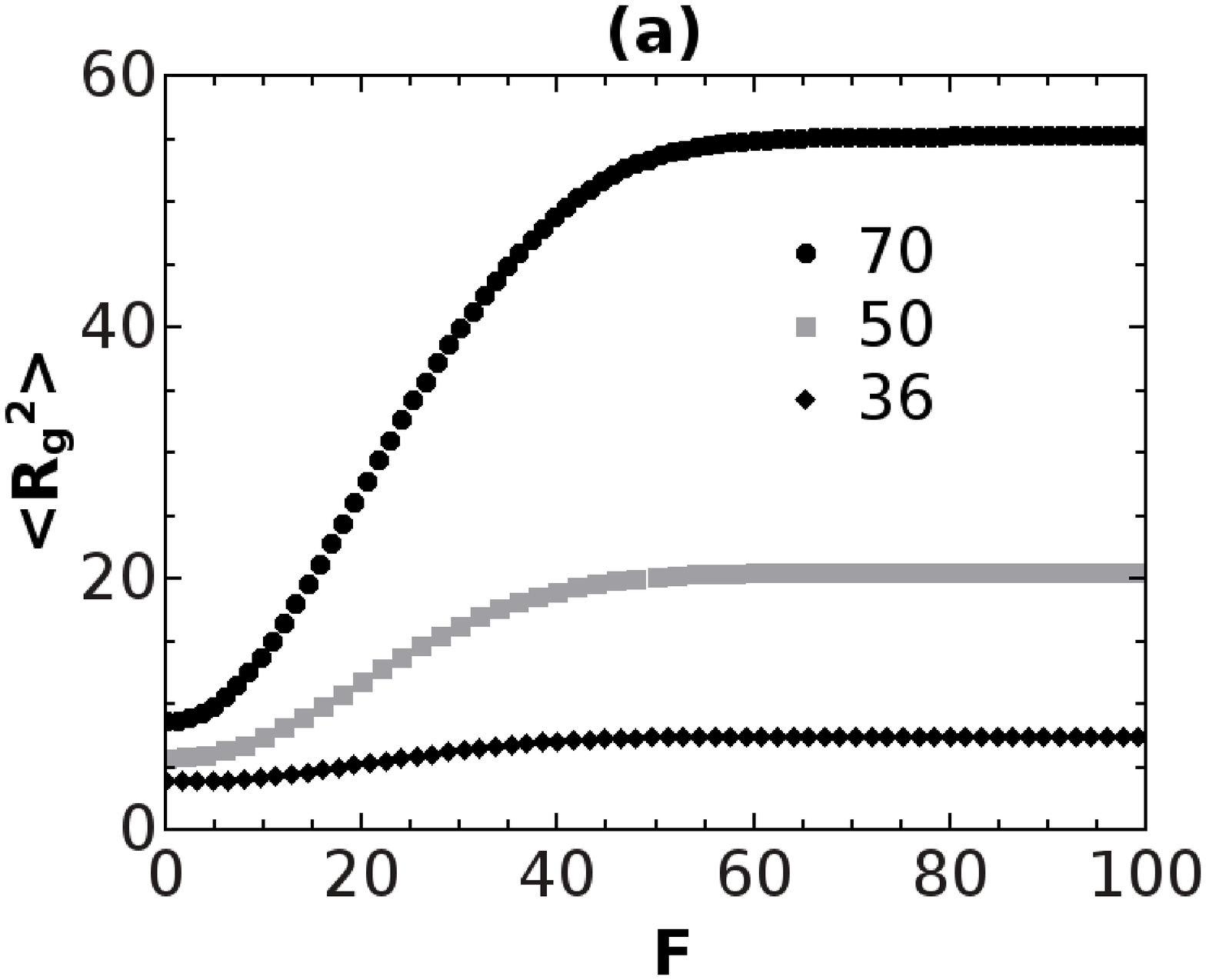}
\includegraphics[width=0.47\textwidth]{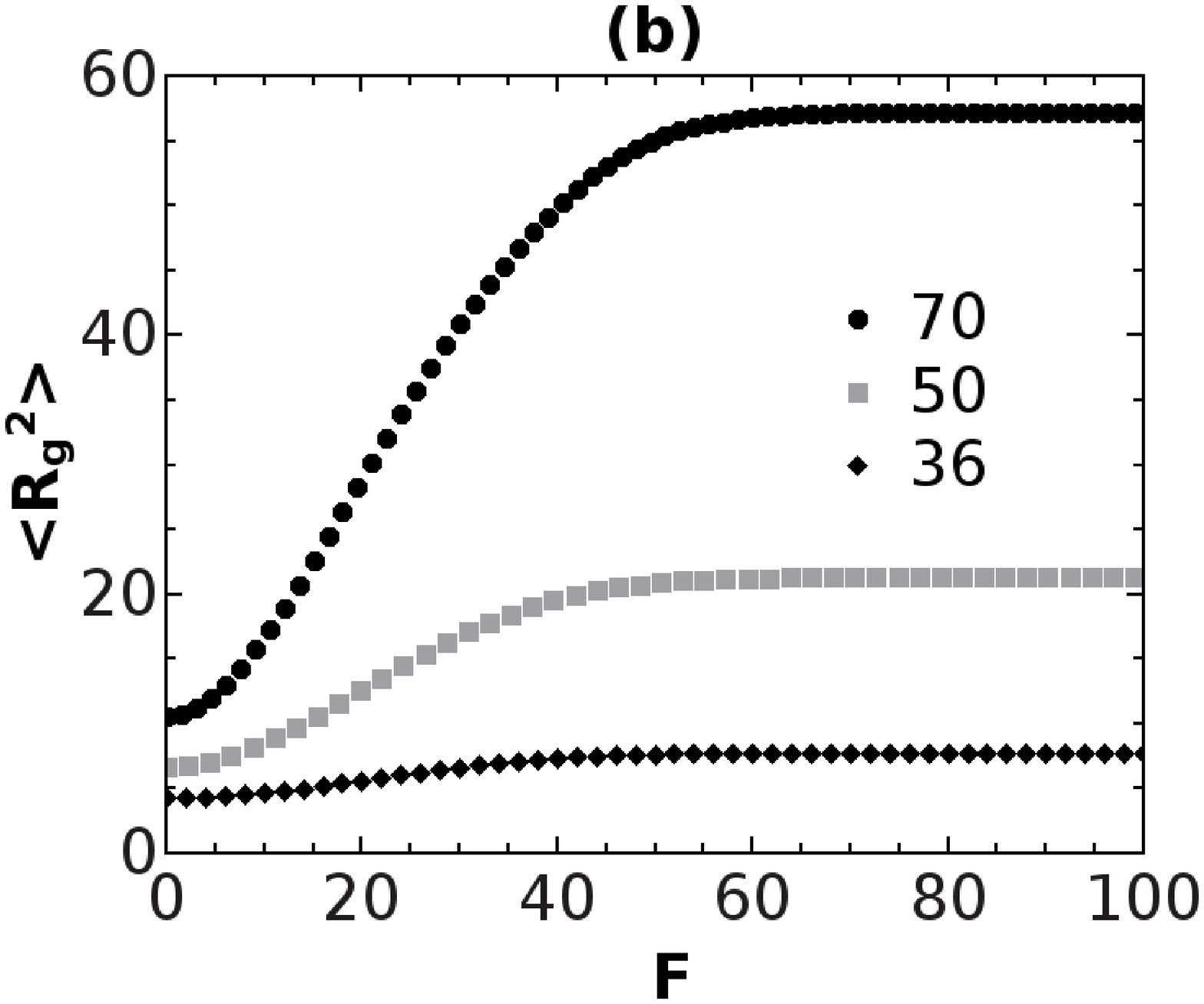}
\caption{The mean square gyration radius of a trefoil knot $3_1$
considered at different lengths $N=36$ (diamonds), $N=50$ (rectangles) and $N=70$
(circles). In (a) the case of attractive interactions ($\varepsilon<0$)
is displayed. In (b) the interactions are repulsive
($\varepsilon>0$). The normalized temperature is  
$\mathbf T=6$.}  \label{rgsizet}   
\end{center} 
\end{figure*}

Regarding the heat capacity, we can see in Fig.~\ref{sizet}~(b)
that there is only one 
peak in the whole studied interval of forces. The interpretation of
this peak is the 
same as in the previous Subsection. In the present
case, the peak  in the heat capacity corresponds to the transition of
the trefoil knot 
from an expanded state to a stretched state. Similar phase 
transitions have already been observed in unknots, see
\cite{swetnam}. It is worth to note that the peak is higher in the case of
longer polymer and its position is slightly shifted toward bigger
values of $\mathbf F$ as
$N$ increases. These are finite size effects that are very well
documented in the polymer literature, see for instance
\cite{finitesizeeffects}. 
Analogous considerations can be made when the interactions are
repulsive.
This case is displayed in
Fig.~\ref{sizet-r}. 
The main difference from knots subjected to attractive interactions
is that in the region in which the
tensile forces are 
very weak, the specific
energy is always positive and not negative as in the case 
$\varepsilon<0$. The reason is that,
for small values of $ F$, the repulsive term $m\varepsilon$, which is
always positive apart from the rare conformations with no contacts at
all, is dominating the 
Hamiltonian (\ref{hx11}). Nonetheless, independently of the sign of $\varepsilon$,
the specific energy of the trefoil knot decreases monotonically in the whole
$\mathbf F$ region, see the insets in
Figs.~\ref{sizet}~(a) when $\varepsilon<0$ and \ref{sizet-r}~(a) when
$\varepsilon>0$. Despite the fact that the
behavior of the heat capacity in Fig.~\ref{sizet-r}~(b)
is very similar to that in which the  interactions are attractive, see
Fig.~\ref{sizet}~(b),
when $\varepsilon<0$, the height of the peak in the latter case is
slightly smaller. This is due to the contribution of the contact term
in the Hamiltonian. Also in the phase transition from crystallites to
expanded states without stretching studied in Refs.~\cite{yzff} and
\cite{yzff2013}, the peak of the heat capacity is higher when the
interactions are attractive.
For the same reason,
the values of $\langle R_G^2\rangle$ at $\varepsilon<0$
are slightly smaller than those measured when $\varepsilon>0$, see
Figs.~\ref{rgsizet}~(a) and (b).
In both cases of $\varepsilon<0$ (Fig.~\ref{rgsizet} (a)) and
$\varepsilon>0$  
(Fig.~\ref{rgsizet} (b)), the gyration radius starts to grow abruptly
more or 
less at the point $\mathbf F=\mathbf F_{peak}$ where the peak in the heat capacity
is appearing.
This similarity of behavior is due to the that we are considering
here the phase transition from 
the expanded to the stretched phase, in which the mechanical term
$-\mathbf Fd_z$ is predominant. In this regime, the repulsive
interactions give a negligible contribution to the average specific energy.
Only when $\mathbf F$ is small, the contact
interactions
become relevant. In a neighborhood of $\mathbf F=0$, indeed,
the values of $\langle
R_g^2 \rangle$ in the case of attractive interactions
are always smaller than those obtained when the monomers repel themselves.

\section{Conclusions}\label{sectionVI}
In this work we have studied the mechanical properties of polymer
knots under stretching by an tensile force $F$ directed along the
$z-$axis. The force has been applied to one point of the knot, while
another point has been anchored to a lattice site. A picture of the
system can be found in Fig.~\ref{force}.
Short polymers with a length up to $N=70$ lattice units have been
considered. In short polymers like these, the effects due to the topology of
the knot are particularly evident. Moreover, for short polymers it is
relatively easy  to sample all the accessible energy states without
the need of introducting cutoffs in the energy range. In performing
the sampling the Wang-Landau algorithm of Ref.~\cite{newwl} has been
proved to be very convenient.
The three dimensional pictures of Figs.~\ref{3t70} (a) and (b) show
the general trend in the behavior of short polymer knots under
stretching. Besides the expanded and crystallite phases, there is
still also a stretched phase. When tensile forces are weak, the knot
can be found in the compact or expanded state depending on the
temperature. The transition from these states to the stretched phase
takes place at any temperature. In the case of a trefoil knot at
high temperatures ($\mathbf T=6$), the
peak of the heat capacity during the transition is located on the
$\mathbf F-$axis at
$\mathbf F_{peak}\sim 35$, see Figs.~\ref{sizet} and \ref{sizet-r}.
With decreasing temperatures
also the values of $\mathbf F_{peak}$ decrease, meaning that it is easier to
stretch a very compact knot at low temperatures than a swollen one at
high temperatures, where strong thermal fluctuations
try to bring back the knot to those conformations that maximize the
entropy of the system.
When the temperature is very low, the stretching process is abrupt and
gives rise to a sharp peak in the heat capacity, see Fig.~\ref{3t70}
(a).
Besides the decreasing of the temperature, there are other factors that
shift $\mathbf F_{peak}$ to lower values, namely the increasing of the
topological complexity of the knot (see Fig.~\ref{top70}~(a)) and the
decreasing of the polymer size (see Figs.~\ref{sizet} and \ref{sizet-r}).
Moreover, almost entirely stretched knots can still pass from a 
compact state to the expanded state with rising temperatures. This
phenomenon causes the appearance of the minor peak in
Fig.~\ref{newfig1}.

Topological effects are more visible when polymers are very short, as
it is proved  by the presence of the double peak of the heat capacity
of the knot $5_1$ with $N=50$ in Fig.~\ref{topology50}~(b).
Surprisingly, these effects fade out rapidly with increasing knot sizes.
For example, the slope of the energy decay of the trefoil knot with
$N=70$ of Fig.~\ref{sizet}~(a) is almost equal to the slope of the
unknot with $N=70$ of
\ref{top70}~(a).
Moreover, in the stretched regime, where the mechanical term becomes
dominant in the Hamiltonian (\ref{hx11}), the behavior of knots
subjected to repulsive or attractive forces is the same apart from
small deviations as it is
expected.

Our work can be generalised under several
aspects. 
First of all, we limited ourselves to the Hamiltonian of
Eq.~(\ref{hx11}) even if there are no obstacles to
to extend our procedure to describe more realistic 
polymer systems, for instance by  
considering the rigidity of polymer knots or adding more complex
potentials considering 
the rigidity of polymer knots.
 This will be
our next research task. We also plan to apply our current method to
proteins which have knotted 
native structures. We hope to be 
able to report on these new developments very soon. 

\begin{acknowledgments} 
The simulations reported in this work were performed in part using the HPC
cluster HAL9000 of the Computing Centre of the Faculty of Mathematics
and Physics at the University of Szczecin.
The work of F. Ferrari results within the collaboration of the COST
Action TD 1308. The use of some of the facilities of the Laboratory of
Polymer Physics of the University of Szczecin, financed by 
a grant of the European Regional Development Fund in the frame of the
project eLBRUS (contract no. WND-RPZP.01.02.02-32-002/10), is
gratefully acknowledged. 
\end{acknowledgments}

\end{document}